\documentclass[prd,superscriptaddress,amsfonts,amssymb,amsmath,showpacs,twocolumn]{revtex4-2}
\usepackage{bm}
\usepackage{amsfonts}
\usepackage{latexsym}
\usepackage[latin1]{inputenc}
\usepackage{graphicx}
\usepackage{amsmath}
\usepackage{palatino}
\usepackage{mathpazo}
\usepackage{textcomp}
\usepackage{float}
\usepackage{booktabs}
\usepackage{dcolumn}
\usepackage{booktabs}
\usepackage{multirow}
\usepackage{hyperref}
\hypersetup{colorlinks,citecolor=blue}
\usepackage{amsmath}
\usepackage{xcolor}
\usepackage{orcidlink}
\usepackage[caption=false]{subfig}
\usepackage{commath}
\def\jnl@style{\it}
\def\aaref@jnl#1{{\jnl@style#1}}

\def\aaref@jnl#1{{\jnl@style#1}}

\def\aj{\aaref@jnl{AJ}}                   
\def\apj{\aaref@jnl{ApJ}}                 
\def\apjl{\aaref@jnl{ApJ}}                
\def\apjs{\aaref@jnl{ApJS}}               
\def\apss{\aaref@jnl{Ap\&SS}}             
\def\aap{\aaref@jnl{A\&A}}                
\def\aapr{\aaref@jnl{A\&A~Rev.}}          
\def\aaps{\aaref@jnl{A\&AS}}              
\def\mnras{\aaref@jnl{Mon.~Not.~Roy.~Astron.~Soc.}}             
\def\prd{\aaref@jnl{Phys.~Rev.~D}}        
\def\prc{\aaref@jnl{Phys.~Rev.~C}}  
\def\prl{\aaref@jnl{Phys.~Rev.~Lett.}}    
\def\qjras{\aaref@jnl{QJRAS}}             
\def\skytel{\aaref@jnl{S\&T}}             
\def\ssr{\aaref@jnl{Space~Sci.~Rev.}}     
\def\zap{\aaref@jnl{ZAp}}                 
\def\nat{\aaref@jnl{Nature}}              
\def\aplett{\aaref@jnl{Astrophys.~Lett.}} 
\def\apspr{\aaref@jnl{Astrophys.~Space~Phys.~Res.}} 
\def\physrep{\aaref@jnl{Phys.~Rep.}}      
\def\physscr{\aaref@jnl{Phys.~Scr}}       
\def\commat{\aaref@jnl{Comm.~Math.~Phys.}}              
\def\science{\aaref@jnl{Science}}               
\def\cqg{\aaref@jnl{Classical Quant.~Grav.}}            
\def\jpcs{\aaref@jnl{JPCS}}                                     
\def\ijmpd{\aaref@jnl{Int.~J.~Mod.~Phys.~D}}                    
\def\grg{\aaref@jnl{Gen.~Relat.~Gravit.}}               
\def\rpp{\aaref@jnl{Rep.~Prog.~Phys.}}          
\def\npa{\aaref@jnl{Nucl.~Phys.~A}}        
\def\lrr{\aaref@jnl{Living Rev.~Rel.}}                   
\def\jcap{\aaref@jnl{J.~Cosmology Astropart.~Phys.}}    
\def\rmp{\aaref@jnl{Rev.~Mod.~Phys.}}   
\def\epjc{\aaref@jnl{Eur.~Phys.~J.~C}}

\allowdisplaybreaks[1]

\begin{document}

\color{black}       

\title{Barrow holographic dark energy models in $f\left( Q\right)$ symmetric
teleparallel gravity with Lambert function distribution}

\author{M. Koussour\orcidlink{0000-0002-4188-0572}}
\email{pr.mouhssine@gmail.com}
\affiliation{Quantum Physics and Magnetism Team, LPMC, Faculty of Science Ben
M'sik,\\
Casablanca Hassan II University,
Morocco.}

\author{S.H. Shekh\orcidlink{0000-0003-4545-1975}}
\email{da\_salim@rediff.com}
\affiliation{Department of Mathematics. S. P. M. Science and Gilani Arts Commerce
College,\\ Ghatanji, Dist. Yavatmal, Maharashtra-445301, India.}

\author{H. Filali}
\email{houda.filali318@gmail.com}
\affiliation{Lab of High Energy Physics, Modeling and Simulations, Faculty of
Science,\\
University Mohammed V-Agdal, Rabat, Morocco.}

\author{M. Bennai\orcidlink{0000-0003-1424-7699}}
\email{mdbennai@yahoo.fr }
\affiliation{Quantum Physics and Magnetism Team, LPMC, Faculty of Science Ben
M'sik,\\
Casablanca Hassan II University,
Morocco.} 
\affiliation{Lab of High Energy Physics, Modeling and Simulations, Faculty of
Science,\\
University Mohammed V-Agdal, Rabat, Morocco.}

\date{\today}

\begin{abstract}
The paper presents Barrow holographic dark energy (infrared cut-off is the
Hubble horizon) suggested by Barrow recently (Physics Letters B 808 (2020):
135643) in an anisotropic Bianchi type-I Universe within the framework of $%
f\left( Q\right) $ symmetric teleparallel gravity, where the non-metricity
scalar $Q$ is responsible for the gravitational interaction. We consider two
cases: Interacting and non-interacting models of pressureless dark matter
and Barrow holographic dark energy by solving $f\left( Q\right) $ symmetric
teleparallel field equations. To find the exact solutions of the field
equations, we assume that the time-redshift relation follows a Lambert
function distribution as $t\left( z\right) =\frac{mt_{0}}{l}g\left( z\right) 
$, where $g\left( z\right) =LambertW\left[ \frac{l}{m}e^{\frac{l-\ln \left(
1+z\right) }{m}}\right] $, $m$ and $l$ are non-negative constants and $t_{0}$
represents the age of the Universe. Moreover, we discuss several
cosmological parameters such as energy density, equation of state (EoS) and
skewness parameters, squared sound speed, and $(\omega _{B}-\omega
_{B}^{^{\prime }})$ plane. Finally, we found the values of the deceleration parameter (DP) for the Lambert function distribution as $q_{(z=0)}=-0.45$ and $q_{(z=-1)}=-1$ which are consistent
with recent observational data, i.e. DP evolves with cosmic time from initial deceleration to late-time acceleration.
\end{abstract}

\maketitle

Astronomical observations from Type Ia supernovae (SNIa), Cosmic Microwave
Background (CMB), Large Scale Structures (LSS), and even more recent data
from multi-wavelength observations of Blazers or the probing of late-time
background expansion using gravitational wave sirens with eLISA have shown
that the Universe is directed towards an accelerated expansion \cite{ref1,
ref2, ref3, ref4, ref5, ref6}. To explain the observed acceleration, dark
energy (DE) was introduced as an added dark component to general relativity
(GR) in the form of the cosmological constant ($\Lambda $) and, with dark
matter (DM), comprises most of the content of the Universe at present time.
Although being the most stable and consistent with observations, the
cosmological constant, which finds its origins in vacuum energy, faces many
constraints mainly the fine-tuning and coincidence problems \cite{ref7}. In
face of these challenges, other forms of dynamical DE that rely on added
exotic forces or matter were proposed such as quintessence, k-essence,
phantom energy, Chaplygin gas, etc \cite{ref8, ref9, ref10, ref11}. 

Another line of research has taken interest in modified gravity theories
(MGT) which were introduced as a set of modifications to Einstein's general
relativity which can describe the accelerated expansion without the need for
an added component or exotic matter. Various theories of modified gravity
were explored in the literature, mainly $f\left( R\right) $ gravity ($R$ is
the Ricci scalar), $f\left( G\right) $ gravity ($G$ is the Gauss-Bonnet
invariant), $f\left( R,T\right) $ gravity ($R$ is the Ricci scalar and $T$
is the trace of the stress-energy tensor), and many extensive scalar-tensor
theories \cite{ref12, ref13, ref14, ref15, ref16, ref17, ref18}. $f\left(
Q\right) $ gravity was introduced as a symmetric teleparallel modification
of gravity and has gained much interest in recent studies as it has shown
many promising results in terms of compatibility with observational
constraints \cite{ref19, ref20, ref21, ref22, ref23, ref24, ref25, ref26,
ref27, ref28, ref29, ref30, ref31, ref32, ref33, ref34, ref35}. It works as
a replacement of geometrical formulations in GR by using a non-metricity
scalar $Q$ as a covariant derivative of the metric tensor. Another
alternative theory that proved its worth is Holographic dark energy, a model
derived from the holographic principle by Susskind et al. that was
introduced to cosmology as a way to probe quantum gravity by assuming that
the entropy of the Universe is proportional to its area \cite{ref36, ref37,
ref38}. By implementing Bekenstein-Hawking black hole thermodynamics and
quantum field theory, Li \cite{ref39}\ introduced a model of dark energy
density constrained by entropy bounds. More recently, Barrow proposed a
modified version of holographic dark energy by taking into account quantum
gravitational effects and fractural features of black holes in the dynamics
of black hole entropy which leads to \cite{ref40}

\begin{equation}
S_{B}=\left( \frac{A}{A_{0}}\right) ^{\frac{\left( 2+\Delta \right) }{2}},
\label{eqn1}
\end{equation}%
where $A$ and $A_{0}$ represent the standard horizon and Planck area,
respectively, and $\Delta $ is a new exponent introduced by Barrow such as $%
0\leq \Delta \leq 1$. For $\Delta =0$, we retrieve the standard
Bekenstein-Hawking entropy. This new form of holographic dark energy has
proven to deliver improved cosmological results compared to its standard
counterpart, see \cite{ref41, ref42, ref43, ref44, ref45}. Motivated by
these attractive results, we explore the effects of Barrow holographic dark
energy (BHDE) with Hubble horizon as the IR cut-off in the background of
anisotropic Bianchi type-I Universe within the framework of $f\left(
Q\right) $ symmetric teleparallel gravity. In reality, the anisotropic
Universe is motivated by Planck's recent results \cite{ref46}, which
confirmed the existence of anomaly in CMB as a result of quantum
fluctuations in the era of cosmic inflation, for more details see \cite%
{ref47}. Moreover, we find the exact solutions of the field equations
assuming the time-redshift relation follows a Lambert function distribution.

This  paper is divided as follows: In Sec. \ref{sec2} we introduce the field
equations of $f\left( Q\right) $ symmetric teleparallel gravity in the
background of anisotropic Bianchi type-I Universe, from which we deduct the
continuity equations of pressureless dark matter and BHDE. In Sec. \ref{sec3}
we establish the solution of the field equations by using cosmological
constraints and the Lambert function distribution. Further, we consider two
cases of study: Interacting and non-interacting $f(Q)$ models which we will
then compare with different existing models of DE. Moreover in Sec. \ref%
{sec4}, we analyze the behavior of the deceleration parameter. Finally in
Sec. \ref{sec5}, we discuss and conclude our results.

\section{Metric and field equations of $f\left( Q\right) $ symmetric
teleparallel gravity}

\label{sec2}

In the present work, we consider the anisotropic LRS Bianchi type-I Universe
metric in the form

\begin{equation}
ds^{2}=-dt^{2}+A^{2}(t)dx^{2}+B^{2}(t)\left( dy^{2}+dz^{2}\right) ,
\label{eqn2}
\end{equation}%
where $A\left( t\right) $ and $B\left( t\right) $ are the metric potentials
of the Universe which are the functions only of the cosmic time $\left(
t\right) $. The Bianchi type-I Universe becomes isotropic if $A\left(
t\right) =B\left( t\right) =a\left( t\right) $. 

Now, we present the basic equations of $f\left( Q\right) $ symmetric
teleparallel gravity. The non-metricity scalar $Q$ is defined as \cite{ref19}

\begin{equation}
Q\equiv -g^{\mu \nu }(L_{\,\,\,\alpha \mu }^{\beta }L_{\,\,\,\nu \beta
}^{\alpha }-L_{\,\,\,\alpha \beta }^{\beta }L_{\,\,\,\mu \nu }^{\alpha }),
\label{eqn3}
\end{equation}%
where the disformation tensor $L_{\alpha \gamma }^{\beta }$ is formulated as,

\begin{equation}
L_{\alpha \gamma }^{\beta }=-\frac{1}{2}g^{\beta \eta }(\nabla _{\gamma
}g_{\alpha \eta }+\nabla _{\alpha }g_{\eta \gamma }-\nabla _{\eta }g_{\alpha
\gamma }).  \label{eqn4}
\end{equation}

The non-metricity tensor is defined in the form

\begin{equation}
Q_{\gamma \mu \nu }=\nabla _{\gamma }g_{\mu \nu },  \label{eqn5}
\end{equation}%
and trace of the non-metricity tensor is derived as follows

\begin{equation}
Q_{\beta }=g^{\mu \nu }Q_{\beta \mu \nu }\qquad \widetilde{Q}_{\beta
}=g^{\mu \nu }Q_{\mu \beta \nu }.  \label{eqn6}
\end{equation}

In addition, we define the superpotential tensor or nonmetricity conjugate
as 

\begin{equation}
P_{\,\,\,\mu \nu }^{\beta }=-\frac{1}{2}L_{\,\,\,\mu \nu }^{\beta }+\frac{1}{%
4}(Q^{\beta }-\widetilde{Q}^{\beta })g_{\mu \nu }-\frac{1}{4}\delta _{(\mu
}^{\beta }Q_{\nu )}.  \label{eqn7}
\end{equation}

From the above equation, the trace of the non-metricity tensor can be
acquired as

\begin{equation}
Q=-Q_{\beta \mu \nu }P^{\beta \mu \nu }.  \label{eqn8}
\end{equation}

The field equations of $f\left( Q\right) $ symmetric teleparallel gravity
are derived from Hilbert--Einstein variational principle. The modified
gravity action is given as%
\begin{equation}
S=\int \left[ \frac{1}{2\kappa }f(Q)+L_{m}\right] d^{4}x\sqrt{-g},
\label{eqn9}
\end{equation}%
where $\kappa =8\pi G$, $f(Q)$ is an arbitrary function of the non-metricity
scalar $Q$, $g$ is the determinant of the metric tensor $g_{\mu \nu }$ i.e. $%
g=\det \left( g_{\mu \nu }\right) $\ and $L_{m}$ is the usual matter
Lagrangian density. The energy-momentum tensor $T_{\mu \nu }$ of matter is
defined as\ 

\begin{equation}
T_{\mu \nu }=\frac{-2}{\sqrt{-g}}\frac{\delta \left( \sqrt{-g}L_{m}\right) }{%
\delta g^{\mu \nu }}.  \label{eqn10}
\end{equation}

Thus, the field equations of $f\left( Q\right) $ symmetric teleparallel
gravity are derived by varying the action $\left( S\right) $ in Eq. (\ref%
{eqn9}) with respect to the metric tensor $g_{\mu \nu }$, 
\begin{widetext}
\begin{equation}
\frac{2}{\sqrt{-g}}\nabla _{\beta }\left( f_{Q}\sqrt{-g}P^{\beta }{}_{\mu
\nu }\right) -\frac{1}{2}fg_{\mu \nu }+f_{Q}\left( P_{\mu \beta \alpha
}Q_{\nu }{}^{\beta \alpha }-2Q^{\beta \alpha }{}_{\mu }P_{\beta \alpha
\nu }\right) =\kappa\left( T_{\mu \nu }+\overline{T}_{\mu \nu }\right) ,
\label{eqn11}
\end{equation}
\end{widetext}where $f_{Q}=\frac{df}{dQ}$, $\nabla _{\beta }$\ is the
covariant derivative, $T_{\mu \nu }$ and $\overline{T}_{\mu \nu }$ are the
energy-momentum tensors of pressureless dark matter and BHDE, respectively.
For simplicity, we use natural units $\left( \kappa =1\right) $.\ In
addition, by varying the action with respect to the connection, we obtain

\begin{equation}
\nabla _{\mu }\nabla _{\beta }\left( f_{Q}\sqrt{-g}P^{\beta }{}_{\mu \nu
}\right) =0.  \label{eqn12}
\end{equation}

The corresponding non-metricity scalar for metric (\ref{eqn2}) can be
written as

\begin{equation}
Q=-2\left( \frac{\overset{.}{B}}{B}\right) ^{2}-4\frac{\overset{.}{A}}{A}%
\frac{\overset{.}{B}}{B}.  \label{eqn13}
\end{equation}

The energy-momentum tensors for pressureless dark matter and BHDE are
defined as 
\begin{widetext}
\begin{equation}
T_{\mu \nu }=\rho _{M}u_{\mu }u_{\nu }=diag\left[ -1,0,0,0\right] \rho _{M},
\label{eqn14}
\end{equation}
\begin{equation}
\overline{T}_{\mu \nu }=\left( p_{B}+\rho _{B}\right) u_{\mu }u_{\nu
}+p_{B}g_{\mu \nu }=diag\left[ -1,\omega _{B},\left( \omega _{B}+\gamma
\right) ,\left( \omega _{B}+\gamma \right) \right] \rho _{B},  \label{eqn15}
\end{equation}
\end{widetext}where, $\rho _{B}$, $\rho _{M}$\ are energy densities of BHDE
and pressureless dark matter, respectively, and $p_{B}$ is the pressure of
BHDE. Here, $\omega _{B}=\frac{p_{B}}{\rho _{B}}$ is the equation of state
(EoS) parameter of the BHDE and $\gamma $ is the deviations from EoS
parameter along $y$ and $z$ directions, known as skewness parameter.

In a commoving co-ordinate system, field equations of $f\left( Q\right) $
symmetric teleparallel gravity (\ref{eqn11}), with Eqs. (\ref{eqn14}) and (%
\ref{eqn15}) for the Bianchi-I Universe (\ref{eqn2}) leads to following
equations of motion \cite{ref28} 
\begin{widetext}
\begin{equation}
\frac{f}{2}+f_{Q}\left[ 4\frac{\overset{.}{A}}{A}\frac{\overset{.}{B}}{B}%
+2\left( \frac{\overset{.}{B}}{B}\right) ^{2}\right] =\rho _{M}+\rho _{B},
\label{eqn16}
\end{equation}
\begin{equation}
\frac{f}{2}-f_{Q}\left[ -2\frac{\overset{.}{A}}{A}\frac{\overset{.}{B}}{B}-2%
\frac{\overset{..}{B}}{B}-2\left( \frac{\overset{.}{B}}{B}\right) ^{2}\right]
+2\frac{\overset{.}{B}}{B}\overset{.}{Q}f_{QQ}=-\rho _{B}\omega _{B},
\label{eqn17}
\end{equation}
\begin{equation}
\frac{f}{2}-f_{Q}\left[ -3\frac{\overset{.}{A}}{A}\frac{\overset{.}{B}}{B}-%
\frac{\overset{..}{A}}{A}-\frac{\overset{..}{B}}{B}-\left( \frac{\overset{.}{%
B}}{B}\right) ^{2}\right] +\left( \frac{\overset{.}{A}}{A}+\frac{\overset{.}{%
B}}{B}\right) \overset{.}{Q}f_{QQ}=-\left( \omega _{B}+\gamma \right) \rho
_{B}.  \label{eqn18}
\end{equation}
\end{widetext}Here, $\left( \text{\textperiodcentered }\right) $ dot
represents a derivative with respect to cosmic time $\left( t\right) $. The
field equations above (\ref{eqn16})-(\ref{eqn18}) can be represented in the
form of mean Hubble and directional Hubble parameters as

\begin{equation}
\frac{f}{2}-Qf_{Q}=\rho _{M}+\rho _{B},  \label{eqn19}
\end{equation}

\begin{equation}
\frac{f}{2}+2\frac{\partial }{\partial t}\left[ H_{y}f_{Q}\right]
+6Hf_{Q}H_{y}=-\rho _{B}\omega _{B},  \label{eqn20}
\end{equation}

\begin{equation}
\frac{f}{2}+\frac{\partial }{\partial t}\left[ f_{Q}\left(
H_{x}+H_{y}\right) \right] +3Hf_{Q}\left( H_{x}+H_{y}\right) =-\left( \omega
_{B}+\gamma \right) \rho _{B},  \label{eqn21}
\end{equation}%
where, we used $\frac{\partial }{\partial t}\left( \frac{\overset{.}{A}}{A}%
\right) =\frac{\overset{..}{A}}{A}-\left( \frac{\overset{.}{A}}{A}\right)
^{2}$ and $Q=-2H_{y}^{2}-4H_{x}H_{y}$. Here, $H=\frac{\overset{.}{a}}{a}=%
\frac{1}{3}\left( H_{x}+2H_{y}\right) $ is the average Hubble parameter and $%
H_{x}=\frac{\overset{.}{A}}{A}$, $H_{y}=H_{z}=\frac{\overset{.}{B}}{B}$
represents the directional Hubble parameters along $x$, $y$ and $z$ axes,
respectively.

Using Eqs. (\ref{eqn19})--(\ref{eqn21}), we obtain the continuity equation
of the pressureless dark matter and BHDE as

\begin{equation}
\overset{.}{\rho }_{M}+\overset{.}{\rho }_{B}+3H\left[ \rho _{M}+\left(
1+\omega _{B}\right) \rho _{B}\right] +2\gamma H_{y}\rho _{B}=0.
\label{eqn22}
\end{equation}%
where the term $\gamma H_{y}\rho _{B}$ in this equation is due to the
consideration of the anisotropic fluid.

\section{Lambert function distribution and cosmological solutions}

\label{sec3}

The above field equations are impossible to find exact solutions to without
adding other constraints, because it is a system consisting of three
independent equations with seven unknowns: $H_{x}$, $H_{y}$, $\rho _{M}$, $%
\rho _{B}$, $\omega _{B}$, $\gamma $ and $f$. There are several constraints
that are used extensively in the literature such as considering the shear
scalar $\left( \sigma ^{2}\right) $ is proportional to the scalar expansion $%
\left( \theta \right) $ i.e.$\ \sigma ^{2}\propto \theta $\ which leads to a
relationship between directional Hubble parameters%
\begin{equation}
H_{x}=kH_{y}\newline
,  \label{eqn23}
\end{equation}%
where $k\neq 1$ is an arbitrary real number which plays a major role in
making the non-isotropic behavior of the Universe. The physical
justification for this condition is imposed on the basis of the observations
of the velocity redshift relation for extragalactic sources which propose
that the Hubble expansion of the Universe may achieve isotropy when $\frac{%
\sigma }{\theta }$ is constant \cite{ref48}. This condition has been used in
many works \cite{ref17, ref28}.

In addition, we assume that the time-redshift relation $t\left( z\right) $
takes the form of a Lambert function distribution as follows%
\begin{equation}
t\left( z\right) =\frac{mt_{0}}{l}g\left( z\right) ,  \label{eqn24}
\end{equation}%
\newline
and%
\begin{equation}
g\left( z\right) =LambertW\left[ \frac{l}{m}e^{\frac{l-\ln \left( 1+z\right) 
}{m}}\right] ,  \label{eqn25}
\end{equation}%
where $m$ and $l$ are non-negative constants and $t_{0}$ represents the age
of the Universe. This time-redshift relation in Eq. (\ref{eqn24}) is
motivated by the hybrid expansion law (HEL) of the scale factor of the
Universe, which is a combination of power law and exponential law i.e. $%
a\left( t\right) =a_{0}\left( \frac{t}{t_{0}}\right) ^{m}e^{l\left( \frac{t}{%
t_{0}}-1\right) }$ where $a_{0}$ represents the present value of scale
factor of the Universe. The HEL of the scale factor of the Universe gives
the exponential law for $m=0$ and the power law for $l=0$ \cite{Akarsu}. 

Using the relation between the mean scale factor and redshift of the
Universe $a\left( t\right) =\left( 1+z\right) ^{-1}$, and taking into
account that the spatial volume $V=AB^{2}$, we find the directional Hubble
parameters as%
\begin{equation}
H_{x}=\frac{3k}{k+2}\left( \frac{m}{t}+\frac{l}{t_{0}}\right) \text{ }\&%
\text{ }H_{y}=H_{z}=\frac{3}{k+2}\left( \frac{m}{t}+\frac{l}{t_{0}}\right) .
\label{eqn26}
\end{equation}

Now, by using the Barrow entropy (\ref{eqn1}), one can obtain the expression
for BHDE energy density as follows

\begin{equation}
\rho _{B}=CL^{\Delta -2},  \label{eqn27}
\end{equation}%
where $C$ is a parameter with dimensions $\left[ L\right] ^{-2-\Delta }$, $L$
can be regarded as the size of the current Universe such as the Hubble scale
and the future event horizon, and $\Delta $ is a free parameter. It can be
seen that the above expression provides the standard holographic dark energy
model $\rho _{\Lambda }=3M_{p}^{2}L^{-2}$ at $\Delta =0$, where $C=3M_{p}^{2}
$ and $c$ the velocity of light equal to unity. In the literature, there are
several possible choices for infrared cut-off $L$ that are found in the
above BHDE density expression. In this work, for simplicity we will assume
the most common form in the literature is the use of Hubble horizon, which
is given as

\begin{equation}
\rho _{B}=CH^{2-\Delta }.  \label{eqn28}
\end{equation}

By using (\ref{eqn26}), the Hubble parameter $\left( H\right) $ for our
cosmological model can be obtained in the form

\begin{equation}
H=\frac{1}{3}\left( H_{x}+2H_{y}\right) =\frac{m}{t}+\frac{l}{t_{0}}.
\label{eqn29}
\end{equation}

Using the above Eqs (\ref{eqn28}) and (\ref{eqn29}), we get the energy
density of the BHDE as

\begin{equation}
\rho _{B}=C\left[ \frac{m}{t}+\frac{l}{t_{0}}\right] ^{2-\Delta }.
\label{eqn30}
\end{equation}

Now, using Eqs. (\ref{eqn23}) and (\ref{eqn26}), we get the non-metricity
scalar in terms of Hubble parameter of this model as

\begin{equation}
Q=\frac{-18\left( 1+2k\right) }{\left( k+2\right) ^{2}}H^{2}.  \label{eqn31}
\end{equation}

We consider the following functional form \cite{ref29, ref49} for our
analysis, which is a combination of a linear and a non-linear term of
non-metricity scalar $Q$,

\begin{equation}
f\left( Q\right) =\alpha Q+\beta Q^{n}.  \label{eqn32}
\end{equation}%
where $\alpha $, $\beta $ and $n\neq 1$ are free model parameters.
Capozziello et al. \cite{Capozziello} found the best approximation for
describing the accelerated expansion of the Universe in $f\left( Q\right) $
gravity is represented by a scenario with $f\left( Q\right) =\alpha +\beta
Q^{n}$. Using Eqs. (\ref{eqn28}), (\ref{eqn31}), (\ref{eqn32}) in (\ref%
{eqn19}), and for this particular $f\left( Q\right) $ cosmological model in
Eq. (\ref{eqn32}), we get the energy density of the pressureless dark matter
in terms of Hubble parameter as 
\begin{widetext}
\begin{equation}
\rho _{M}=\frac{9\alpha \left( 1+2k\right) }{\left( k+2\right) ^{2}}H^{2}+%
\frac{\left( -18\right) ^{n}\beta \left( 1-2n\right) \left( 1+2k\right) ^{n}%
}{2\left( k+2\right) ^{2n}}H^{2n}-CH^{2-\Delta }.  \label{eqn33}
\end{equation}
\end{widetext}Using Eqs. (\ref{eqn19}), (\ref{eqn20}), (\ref{eqn28}), (\ref%
{eqn31}), (\ref{eqn32}) in (\ref{eqn21}) we get the skewness parameter as

\begin{equation}
\gamma =\frac{\gamma _{1}\left( \overset{.}{H}+3H^{2}\right) }{H^{2-\Delta }}%
+\gamma _{2}\left[ \left( 2n-1\right) \frac{\overset{.}{H}}{H^{2}}+3\right] 
\frac{H^{2n}}{H^{2-\Delta }},  \label{eqn34}
\end{equation}%
where

\begin{equation}
\gamma _{1}=\frac{3\alpha \left( 1-k\right) }{C\left( k+2\right) },
\label{eqn35}
\end{equation}%
and

\begin{equation}
\gamma _{2}=\frac{3\beta n\left( -18\right) ^{n-1}\left( 1-k\right) \left(
1+2k\right) ^{n-1}}{C\left( k+2\right) ^{2n-1}}.  \label{eqn36}
\end{equation}%

\begin{figure}[H]
\centering
\includegraphics[scale=0.6]{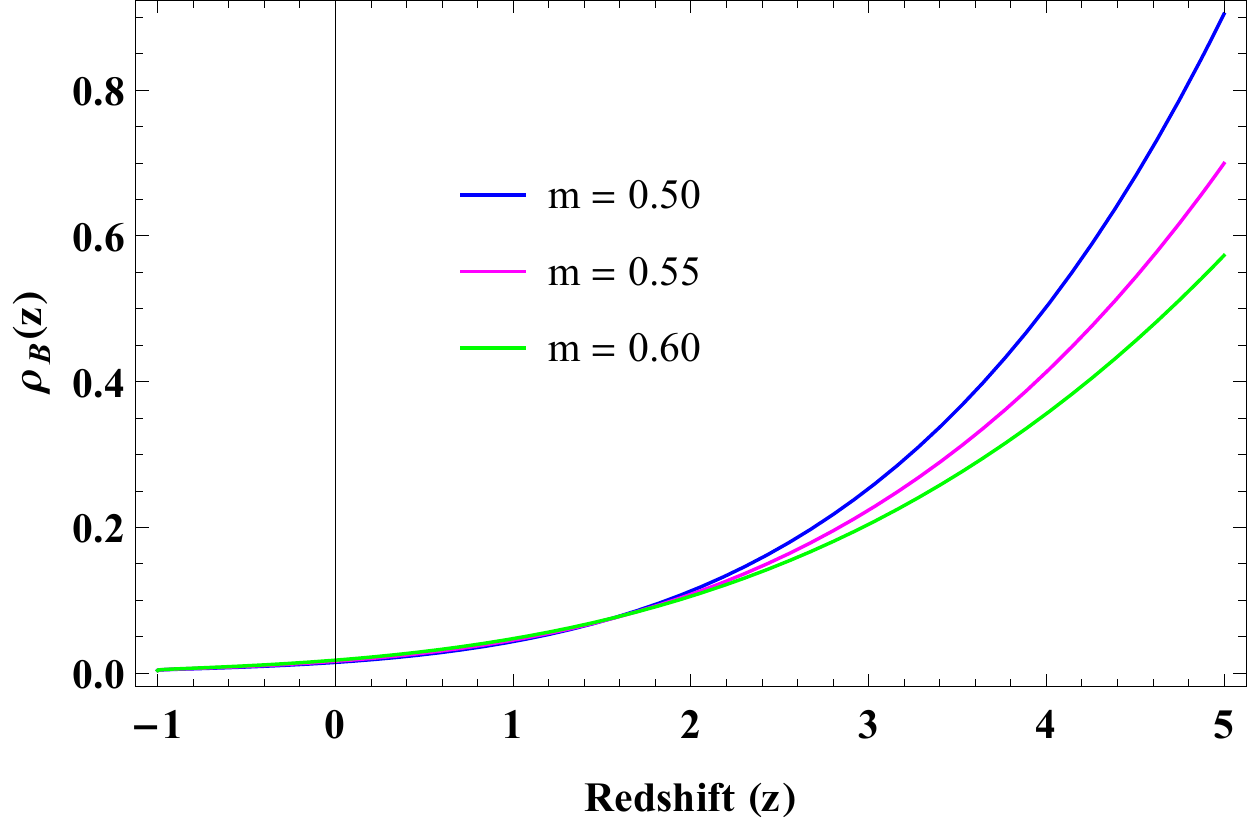}
\caption{Plot of energy density $\left( \protect\rho _{B}\right) $ of BHDE
vs. redshift $\left( z\right) $ for $\protect\alpha =1$, $\protect\beta =-1$%
, $n=C=2$, $\Delta =0.2$, $l=0.4$, and $t_{0}=13.8$.}
\label{fig1}
\end{figure}

\begin{figure}[H]
\centering
\includegraphics[scale=0.6]{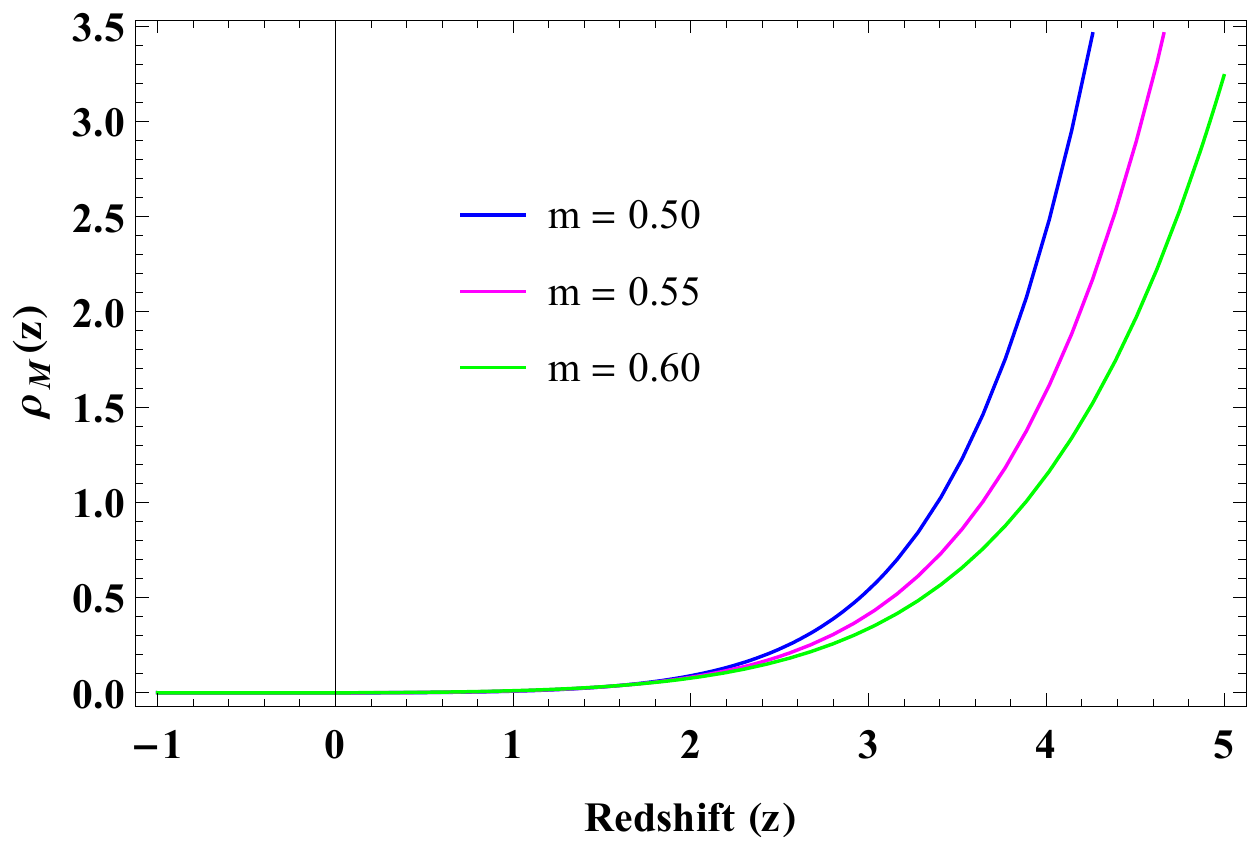}
\caption{Plot of energy density $\left( \protect\rho _{M}\right) $ of
matter vs. redshift $\left( z\right) $ for $\protect\alpha =1$, $\protect%
\beta =-1$, $n=C=2$, $\Delta =0.2$, $l=0.4$, and $t_{0}=13.8$.}
\label{fig2}
\end{figure}

\begin{figure}[H]
\centering
\includegraphics[scale=0.6]{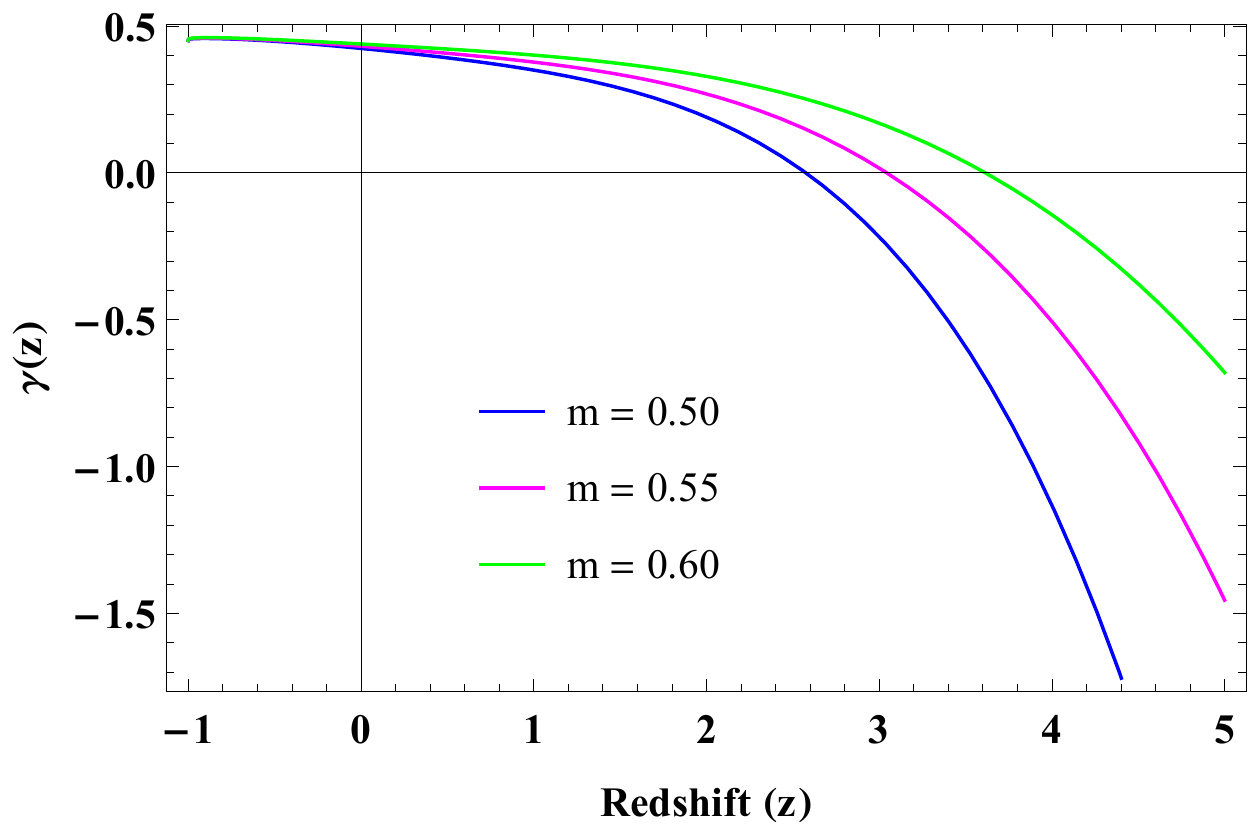}
\caption{Plot of skewness parameter $\left( \protect\gamma \right) $ vs.
redshift $\left( z\right) $ for $\protect\alpha =1$, $\protect\beta =-1$, $%
n=C=2$, $\Delta =0.2$, $l=0.4$, and $t_{0}=13.8$.}
\label{fig3}
\end{figure}

In Figs. \ref{fig1} and \ref{fig2} we have plotted the behaviors of
pressureless dark matter density $\left( \rho _{M}\right) $ and BHDE density 
$\left( \rho _{B}\right) $ with the Hubble horizon cut-off in terms of
redshift $\left( z\right) $ for the three different values of $m=0.50$, $0.55
$, $0.60$, respectively. We can see that both $\rho _{M}$ and $\rho _{B}$
are increasing functions with redshift and positive for all $z$ values.
Moreover, Fig. \ref{fig3} represents the behavior of skewness parameter $%
\left( \gamma \right) $\ in terms of redshift $\left( z\right) $ for the
three different values of $m$. From the figure, it is clear that $\gamma $
is positive at the initial time, and negative at the present i.e. $%
z\rightarrow 0$ and future i.e. $z\rightarrow -1$. Hence, the BHDE $f\left(
Q\right) $\ model is anisotropic throughout evolution of the Universe. In
the following sections, we will discuss two cases: Non-interacting and
interacting $f\left( Q\right) $\ model. In addition, we compare these two
cases with models of DE in the literature such as the quintessence, phantom, 
$\Lambda CDM$, etc.

\subsection{Phantom like behavior of $f\left( Q\right) $\ non-interacting
model}

In this subsection, we consider that there is no energy exchange between the
two basic components of the Universe: the pressureless dark matter component
and BHDE component. Therefore, the continuity equation (\ref{eqn22}) can be
written as 
\begin{equation}
\overset{.}{\rho }_{M}+3H\rho _{M}=0.  \label{eqn37}
\end{equation}%
\begin{equation}
\overset{.}{\rho }_{B}+3H\left( 1+\omega _{B}\right) \rho _{B}+2\gamma
H_{y}\rho _{B}=0.  \label{eqn38}
\end{equation}

Using Eqs. (\ref{eqn28}) and (\ref{eqn34}) in Eq. (\ref{eqn38}), we get the
EoS parameter of BHDE as 
\begin{equation}
\omega _{B}=-1-\left[ \frac{\left( 2-\Delta \right) }{3}\frac{\overset{.}{H}%
}{H^{2}}+\frac{2\gamma }{\left( k+2\right) }\right] .  \label{eqn39}
\end{equation}%

\begin{figure}[H]
\centering
\includegraphics[scale=0.6]{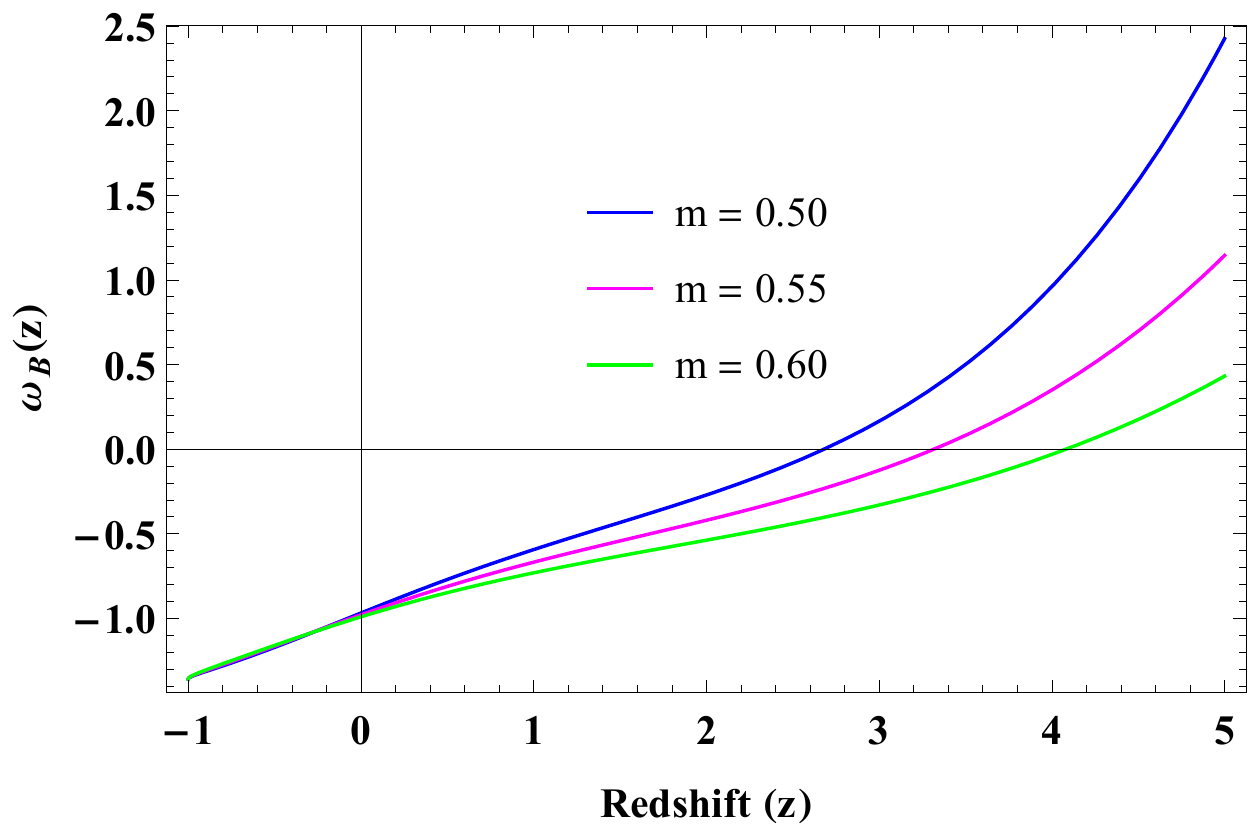}
\caption{Plot of EoS parameter $\left( \protect\omega _{B}\right) $ vs.
redshift $\left( z\right) $ for $\protect\alpha =1$, $\protect\beta =-1$, $%
n=C=2$, $\Delta =0.2$, $l=0.4$, and $t_{0}=13.8$.}
\label{fig4}
\end{figure}

\begin{figure}[H]
\centering
\includegraphics[scale=0.6]{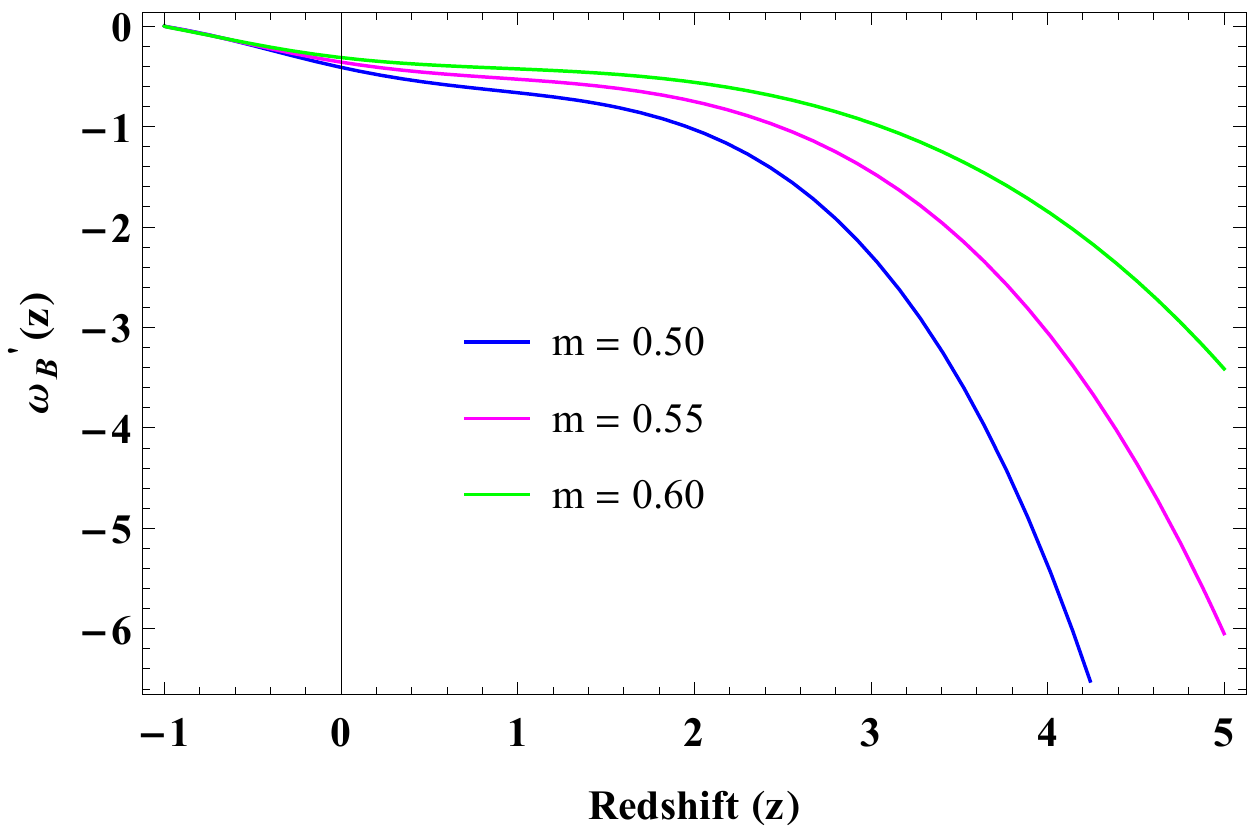}
\caption{Plot of $\protect\omega _{B}^{^{\prime }}$ vs. redshift $\left(
z\right) $ for $\protect\alpha =1$, $\protect\beta =-1$, $n=C=2$, $\Delta
=0.2$, $l=0.4$, and $t_{0}=13.8$.}
\label{fig6}
\end{figure}

In this work \cite{ref50} Caldwell and Linder separate the quintessence
phase of DE into two different regions: thawing ($\omega _{B}^{^{\prime }}>0,
$ $\omega _{B}<0$) and freezing ($\omega _{B}^{^{\prime }}<0,$ $\omega _{B}<0
$) regions by introducing a new analysis called $\omega _{B}-\omega
_{B}^{^{\prime }}$ plane. For $\omega _{B}^{^{\prime }}$ prime designate the
derivative of EoS parameter with respect to $x=\ln a$. Using Eq. (\ref{eqn39}%
), we get

\begin{equation}
\omega _{B}^{^{\prime }}=\frac{1}{H}\left[ \frac{\left( 2-\Delta \right) }{3}%
\left\{ 2\frac{\overset{.}{H}^{2}}{H^{3}}-\frac{\overset{..}{H}}{H^{2}}%
\right\} -\frac{2\overset{.}{\gamma }}{\left( k+2\right) }\right] ,
\label{eqn40}
\end{equation}%
where $\overset{.}{\gamma }=H^{\Delta -5}\left[ 
\begin{array}{c}
\text{$\gamma _{2}$}(2n-1)H^{2n+1}\overset{..}{H} \\ 
+\text{$\gamma $}_{1}H^{3}\overset{..}{H}+\text{$\gamma $}_{2}(2n-1)(\Delta
+2n-4) \\ 
\times H^{2n}\overset{.}{H}^{2} \\ 
+3\text{$\gamma $}_{2}(\Delta +2n-2)H^{2n+2}\overset{.}{H}+ \\ 
3\text{$\gamma $}_{1}\Delta H^{4}\overset{.}{H}+\text{$\gamma $}_{1}(\Delta
-2)H^{2}\overset{.}{H}^{2}%
\end{array}%
\right] $,\newline
$\overset{.}{H}=-\frac{m}{t^{2}}$ and $\overset{..}{H}=\frac{2m}{t^{3}}$. In
this background, the squared sound speed ($v_{s}^{2}$) is exploited for
examining the stability of the dark energy models which is explicit as $%
v_{s}^{2}=\frac{dp_{B}}{d\rho _{B}}=\frac{\overset{.}{p}_{B}}{\overset{.}{%
\rho }_{B}}$. If $v_{s}^{2}>0$, we obtain a stable model and if $v_{s}^{2}<0$%
, we obtain unstable model. For our non-interacting BHDE $f\left( Q\right) $%
\ model $v_{s}^{2}$\ takes the following form

\begin{eqnarray}
v_{s}^{2} &=&-1+\frac{1}{3}\left[ \left\{ -\frac{(2-\Delta )\overset{.}{H}}{%
H^{2}}-\frac{6\gamma }{\left( k+2\right) }\right\} +\frac{H}{(2-\Delta )%
\overset{.}{H}}\right]   \label{eqn41} \\
&&\times \frac{1}{3}\left[ \left\{ -\frac{(2-\Delta )\overset{..}{H}}{H^{2}}+%
\frac{2(2-\Delta )\overset{.}{H}^{2}}{H^{3}}-\frac{6\overset{.}{\gamma }}{%
\left( k+2\right) }\right\} \right] .  \notag
\end{eqnarray}

\begin{figure}[H]
\centering
\includegraphics[scale=0.6]{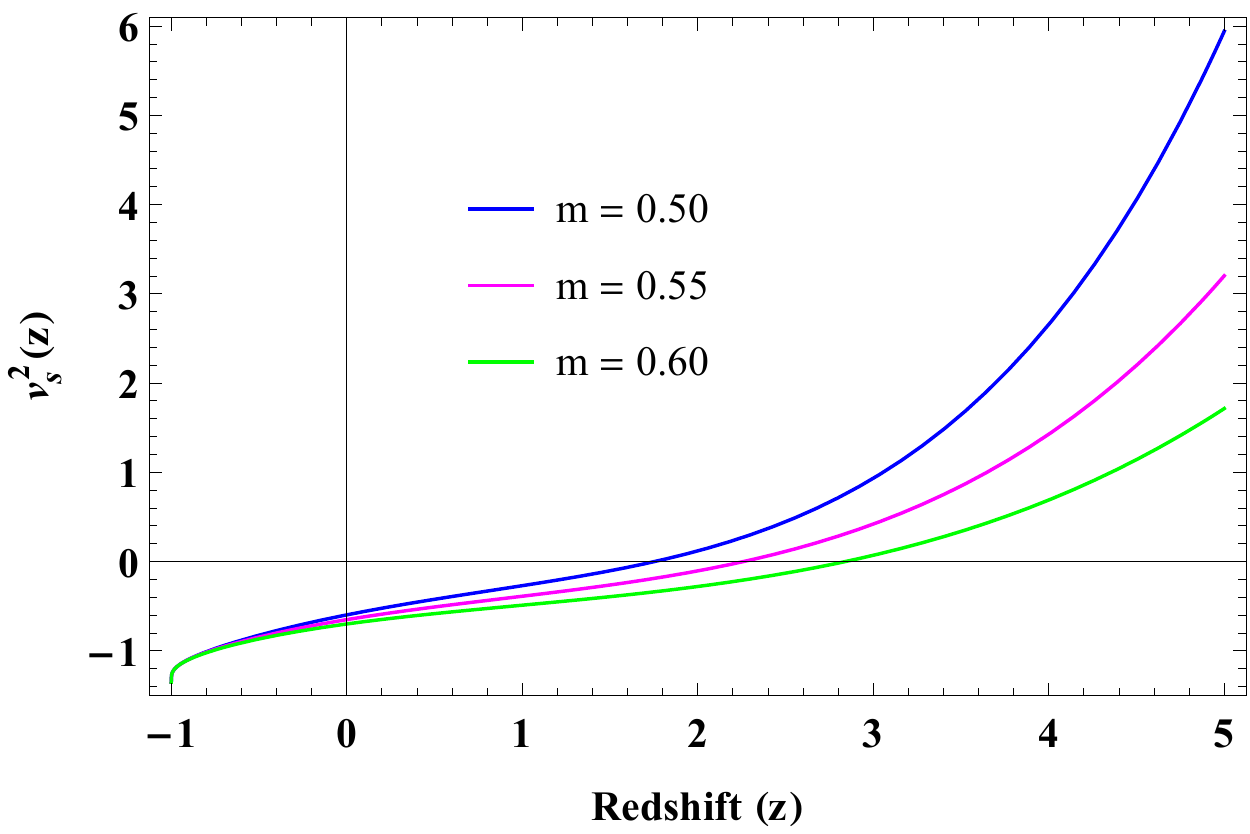}
\caption{Plot of squared sound speed $\left( v_{s}^{2}\right) $ vs.
redshift $\left( z\right) $ for $\protect\alpha =1$, $\protect\beta =-1$, $%
n=C=2$, $\Delta =0.2$, $l=0.4$, and $t_{0}=13.8$.}
\label{fig7}
\end{figure}

In Fig. \ref{fig4} we plot the behavior of the EoS parameter
($\omega _{B}$) of non-interacting BHDE $f\left( Q\right) $\ model in terms
of redshift ($z$) for three different values of $m=0.50$, $0.55$, $0.60$.
These results can be interpreted as follows: At the beginning of time, the
EoS parameter starts from the matter-dominated era, then it moves to the
quintessence region ($-1<\omega _{B}<-0.33$) and crosses the $\Lambda $CDM
model ($\omega _{B}=-1$) in the current time and finally approaches to a
phantom region ($\omega _{B}<-1$). Further, the current values of the EoS
parameter are $\omega _{B}\sim -1$ ($z=0$) for the three values of $m$.
Thus, these values are consistent with Planck 2018 data \cite{ref51}. The $%
\omega _{B}^{^{\prime }}$ parameter for non-interacting BHDE $f\left(
Q\right) $\ model for three different values of $m$ versus redshift $(z)$ is
plotted in Fig. \ref{fig6}. It is clear from Figs. \ref{fig4} and \ref{fig6}
that the $\omega _{B}-\omega _{B}^{^{\prime }}$ plane corresponds to
freezing region for three different values of $m$.\ Fig. \ref{fig7} shows
the evolution of the squared sound speed ($v_{s}^{2}$) in terms of redshift (%
$z$). We can see that $v_{s}^{2}$ is positive in the initial time i.e. our
model is stable, and negative in the present and future i.e. an unstable
model.

\subsection{$\Lambda $CDM like behavior of $f\left( Q\right) $\ interacting
model}

In this case, we assume that the pressureless dark matter component is
interacting with the BHDE component via the interaction term $Q$, we can
write the continuity equation of pressureless dark matter and BHDE as

\begin{equation}
\overset{.}{\rho }_{M}+3H\rho _{M}=Q.  \label{eqn42}
\end{equation}

\begin{equation}
\overset{.}{\rho }_{B}+3H\left( 1+\omega _{B}\right) \rho _{B}+2\gamma
H_{y}\rho _{B}=-Q.  \label{eqn43}
\end{equation}

From the above continuity equation, we can see that the interaction term
must be proportional to a quantity with units of inverse of cosmic time.
Therefore, this term in the literature can take several forms ($Q$-classes)
such as $Q=3\eta H\rho _{M}$, $Q=3\eta H\rho _{DE}$, and $Q=3\eta H\left(
\rho _{M}+\rho _{DE}\right) $ \cite{ref52, ref53, ref54}. In this study, we
choose $Q=3\eta H\rho _{B}$ as an interaction term where $3\eta H$ is the
decay rate with a coupling constant $\eta $ (interaction parameter) \cite%
{ref55}. In general, the interaction parameter $\eta $ can be positive or
negative. If $\eta $ is positive means BHDE decays to pressureless DM, while
if $\eta $ is negative means pressureless DM decays to BHDE. The previous
situation of the non-interacting $f\left( Q\right) $ model can be obtained
with $\eta =0$.

Using Eqs. (\ref{eqn28}) and (\ref{eqn34}) in (\ref{eqn43}), we get the EoS
parameter for this case as

\begin{equation}
\omega _{B}=-1-\eta -\left[ \frac{\left( 2-\Delta \right) }{3}\frac{\overset{%
.}{H}}{H^{2}}+\frac{2\gamma }{\left( k+2\right) }\right] .  \label{eqn44}
\end{equation}

Using the same method in the previous case, we find the derivative of $%
\omega _{B}$ with respect to $x=\ln a$ as follows

\begin{equation}
\omega _{B}^{^{\prime }}=\frac{1}{H}\left[ \frac{\left( 2-\Delta \right) }{3}%
\left\{ 2\frac{\overset{.}{H}^{2}}{H^{3}}-\frac{\overset{..}{H}}{H^{2}}%
\right\} -\frac{2\overset{.}{\gamma }}{\left( k+2\right) }\right] .
\label{eqn45}
\end{equation}

The squared sound speed ($v_{s}^{2}$) in this case is derived as 
\begin{widetext}
\begin{equation}
v_{s}^{2}=-1-\eta +\frac{1}{3}\left[ \frac{(\Delta -2)\overset{.}{H}}{H^{2}}-%
\frac{H}{(\Delta -2)\overset{.}{H}}\left\{ \frac{(\Delta -2)\overset{..}{H}}{%
H^{2}}-\frac{2(\Delta -2)\overset{.}{H}^{2}}{H^{3}}-\frac{6\overset{.}{%
\gamma }}{\left( k+2\right) }\right\} -\frac{6\gamma }{\left( k+2\right) }%
\right] .  \label{eqn46}
\end{equation}
\end{widetext}

\begin{figure}[H]
\centering
\includegraphics[scale=0.6]{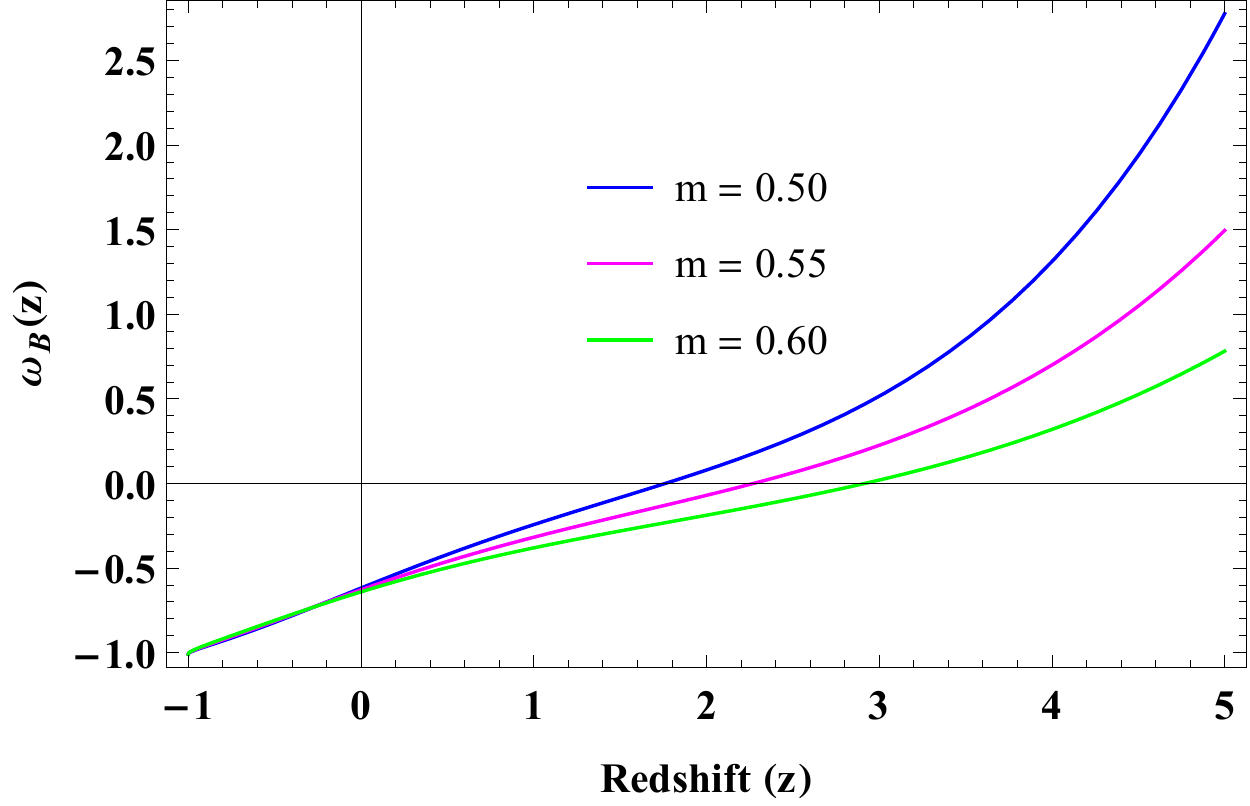}
\caption{Plot of EoS parameter $\left( \protect\omega _{B}\right) $ vs.
redshift $\left( z\right) $ for $\protect\alpha =1$, $\protect\beta =-1$, $%
n=C=2$, $\Delta =0.2$, $l=0.4$, $t_{0}=13.8$ and $\protect\eta=-0.35$.}
\label{fig8}
\end{figure}

\begin{figure}[H]
\centering
\includegraphics[scale=0.6]{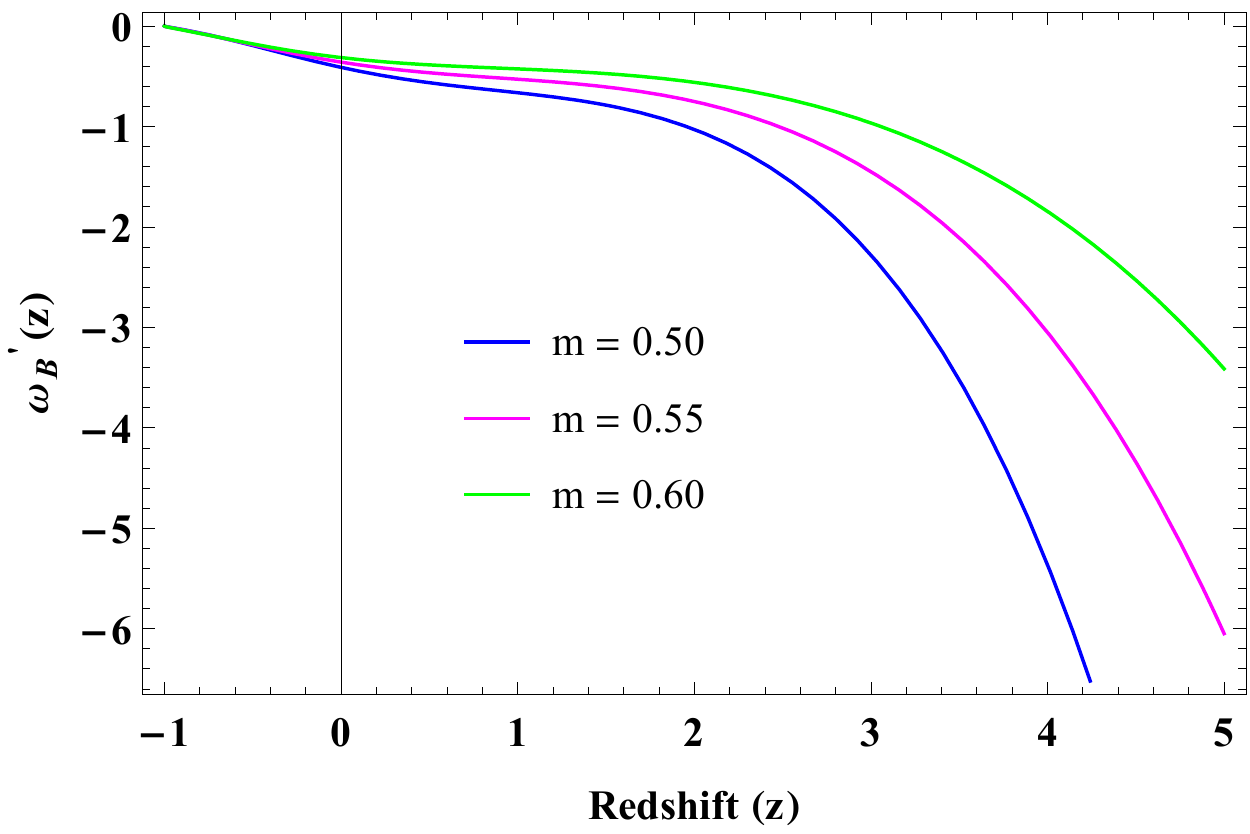}
\caption{Plot of $\protect\omega _{B}^{^{\prime
}} $ vs. redshift $\left( z\right) $ for $\protect\alpha =1$, $\protect\beta =-1$, $n=C=2$, $\Delta =0.2$, $%
l=0.4$, $t_{0}=13.8$ and $\protect\eta=-0.35$.}
\label{fig10}
\end{figure}

\begin{figure}[H]
\centering
\includegraphics[scale=0.6]{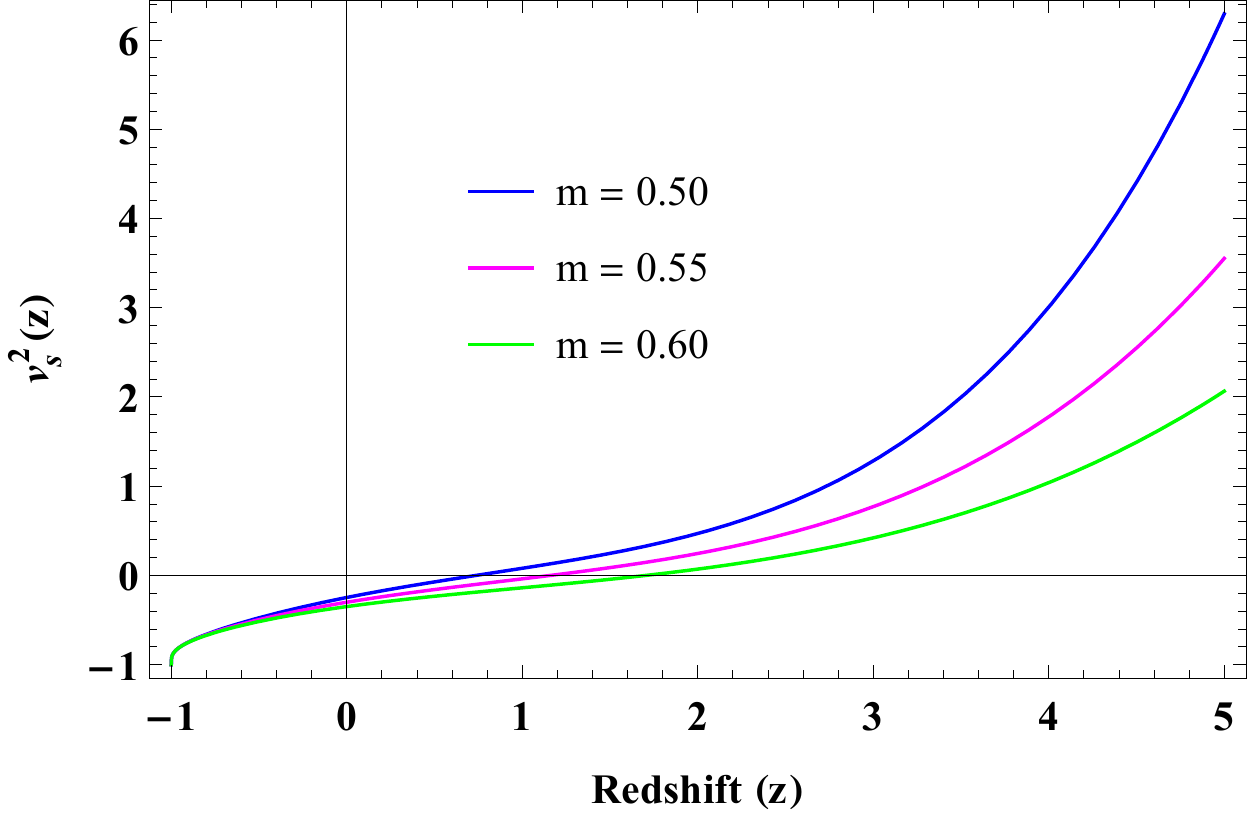}
\caption{Plot of squared sound speed $\left( v_{s}^{2}\right) $ vs.
redshift $\left( z\right) $ for $\protect\alpha =1$, $\protect\beta =-1$, $%
n=C=2$, $\Delta =0.2$, $l=0.4$, $t_{0}=13.8$ and $\protect\eta=-0.35$.}
\label{fig11}
\end{figure}

Fig. \ref{fig8} describes the behavior of EoS parameter ($%
\omega _{B}$) for interacting BHDE $f\left( Q\right) $ model in terms of
redshift ($z$) for three different values of $m$. We also observe that the
model begins from a matter-dominated era, varies in the quintessence region,
and finally approaches to standard $\Lambda CDM$ model. Further, the current
value of $\omega _{B}$ corresponds to the most recent data. The $\omega
_{B}^{^{\prime }}$ parameter for interacting BHDE $f\left( Q\right) $\ model
versus redshift $(z)$ for three different values of $m$ is plotted in Fig. %
\ref{fig10}. It is clear that the $\omega _{B}-\omega _{B}^{^{\prime }}$
plane corresponds to freezing region for three different values of $m$. Fig. %
\ref{fig11} shows the evolution of the squared sound speed ($v_{s}^{2}$)
versus redshift ($z$). It can be observed that $v_{s}^{2}$ of interacting
BHDE $f\left( Q\right) $ model is positive in the initial time i.e. the
model is stable, and negative in the present and future i.e. we get an
unstable model.

\section{Deceleration parameter}

\label{sec4}

To verify that the proposed model predicts an accelerating phase of the
Universe, we study the behavior of the deceleration parameter (DP) of our
cosmological models. The DP sign indicates if the model is accelerating or
decelerating. If $q>0$, the model with a deceleration expansion, if $q=0$ a
constant rate of expansion and an accelerated expansion if $q<0$. The DP for
our cosmological models is given by

\begin{equation}
q=-1+\frac{d}{dt}\left( \frac{1}{H}\right) =-1+mt_{0}^{2}\left(
mt_{0}+lt\right) ^{-2}  \label{eqn47}
\end{equation}

\begin{figure}[H]
\centering
\includegraphics[scale=0.6]{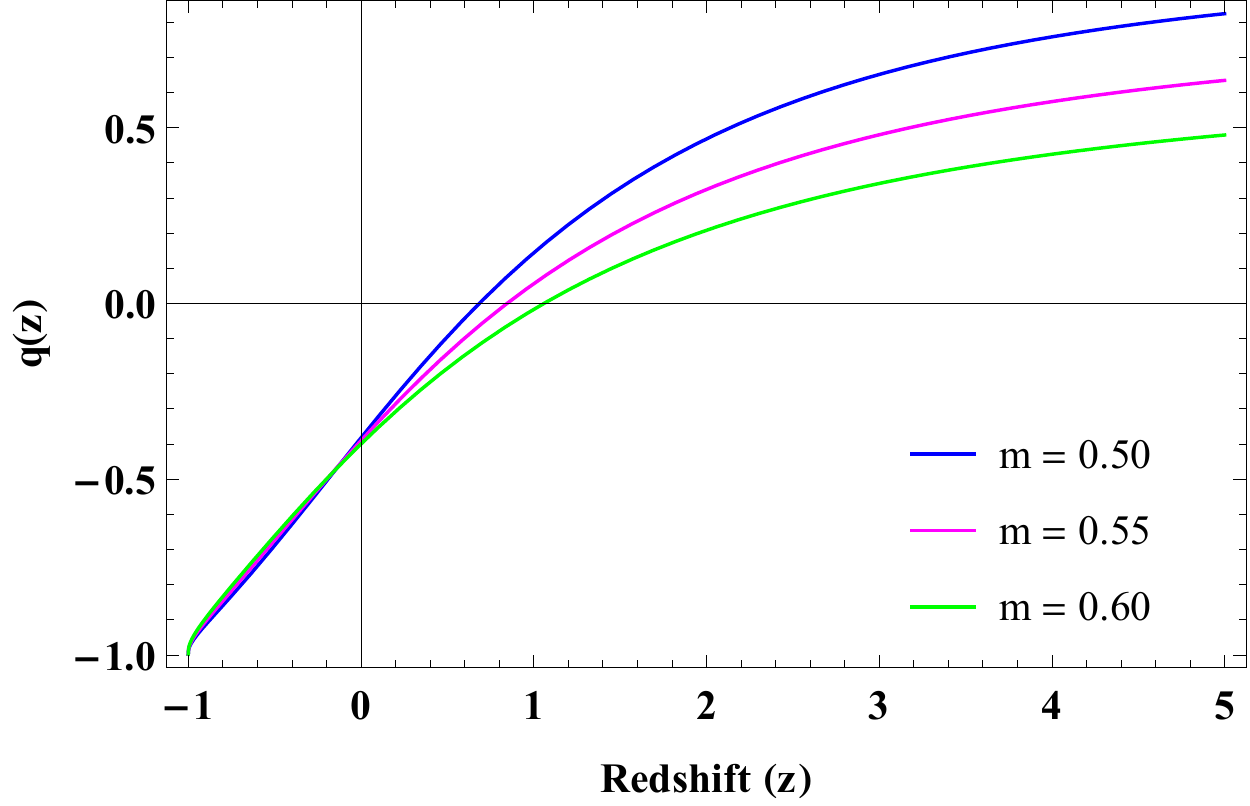}
\caption{Plot of deceleration parameter $q$ vs. redshift $z$ for $\alpha =1$, $\beta =-1$, $n=C=2$, $\Delta =0.2$, $%
l=0.4$, $t_{0}=13.8$ and $\eta=-0.35$.}
\label{fig12}
\end{figure}

The behavior of DP ($q$) in terms of redshift ($z$) is shown in Fig. \ref{fig12}. It can be seen that the DP for our models evolves
with cosmic time from initial deceleration with positive values to late-time
acceleration with negative values and finally approaches to $-1$. Further,
the current values $q_{0}\left( z=0\right) $\ of the DP correspond to the
observational data of SNeIa and CMBR.

\section{Conclusions}

\label{sec5}

In this paper, we have investigated the Barrow holographic dark energy in an
anisotropic Bianchi type-I Universe within the framework of $f\left(
Q\right) $ symmetric teleparallel gravity, where the non-metricity scalar $Q$
is responsible for the gravitational interaction. To discuss the current
cosmic acceleration, we considered two cases for the study: Interacting and
non-interacting models of pressureless dark matter and BHDE. Then we used
two main hypotheses in this work: (i) we assumed that the shear scalar $%
\left( \sigma ^{2}\right) $ is proportional to the scalar expansion $\left(
\theta \right) $ i.e.$\ \sigma ^{2}\propto \theta $\ which leads to a
relationship between directional Hubble parameters as $H_{x}=kH_{y}$, where $%
k\neq 0,1$, (ii) we assumed that the redshift-time relation follows the form
of a Lambert function distribution. In addition, we considered the $f\left(
Q\right) $ model as a combination of a linear and a non-linear term of
non-metricity scalar $Q$ i.e. $f\left( Q\right) =\alpha Q+\beta Q^{n}$,
where $\alpha $, $\beta $ and $n\neq 1$ are free model parameters. We have
discussed the behavior of various cosmological parameters that are used in
this context, and the following are the most important results obtained: We
observed for our models that both pressureless dark matter density and BHDE
density are increasing functions with redshift and positive for all $z$
values (Figs. \ref{fig1} and \ref{fig2}). Further, we observed that the
skewness parameter is positive at the initial time and negative at the
present and future (Fig. \ref{fig3}). Hence, the BHDE $f\left( Q\right) $\
model is anisotropic throughout evolution of the Universe. 

Another interesting result of our cosmological models is that the EoS
parameter of non-interacting BHDE $f\left( Q\right) $ model is similar to
phantom model and interacting BHDE $f\left( Q\right) $ model like $\Lambda
CDM$ (Figs. \ref{fig4} and \ref{fig8}). The
evolution of the $\omega _{B}-\omega _{B}^{^{\prime }}$ plane for both
models: Non-interacting and interacting BHDE $f\left( Q\right) $\ models
corresponds to freezing region ($\omega _{B}^{^{\prime }}<0,\omega _{B}<0$)
for three different values of $m$. Furthermore, we investigated the behavior
of squared sound speed for both models. We also found that both models are
stable at the beginning of time and unstable at the present and future
periods. Finally, the evolution of the deceleration parameter in Fig. \ref{fig12} indicates a transition of the Universe from
decelerated to accelerated phase. Further, we found the values of the deceleration parameter (DP) for the Lambert function distribution as $q_{(z=0)}=-0.45$ and $q_{(z=-1)}=-1$ which are consistent
with recent observational data.

\section*{Acknowledgments}

We are very much grateful to the honorary referee and the editor for the
illuminating suggestions that have significantly improved our work in terms
of research quality and presentation.\newline


\begin{thebibliography}{99}
\bibitem{ref1} A. G. Riess et al., Observational evidence from
supernovae for an accelerating Universe and a cosmological constant, \textit{The
Astronomical Journal} \textbf{116} (1998) 1009.

\bibitem{ref2} A. G. Riess et al., Type Ia supernova discoveries at $z>1$
from the Hubble Space Telescope: Evidence for past deceleration and
constraints on dark energy evolution. \textit{The Astrophysical Journal} \textbf{607} (2004) 665.

\bibitem{ref3} S. Hanany et al., MAXIMA-1: a measurement of the cosmic
microwave background anisotropy on angular scales of 10'-5, \textit{The
Astrophysical Journal} \textbf{545} (2000) L5.

\bibitem{ref4} A. Dom\'{\i}nguez and P. Francisco, Measurement of
the Expansion Rate of the Universe from $U{3b3}$-Ray Attenuation, \textit{The
Astrophysical Journal Letters} \textbf{771} (2013) L34.

\bibitem{ref5} D. J. Eisenstein et al., Detection of the baryon
acoustic peak in the large-scale correlation function of SDSS luminous red
galaxies. \textit{The Astrophysical Journal} \textbf{633} (2005) 560.

\bibitem{ref6} N. Tamanini et al., Science with the space-based
interferometer eLISA. III: Probing the expansion of the Universe using
gravitational wave standard sirens, \textit{Journal of Cosmology and Astroparticle
Physics}. \textbf{04} (2016) 002.

\bibitem{ref7} C. Armendariz-Picon, V. Mukhanov, and S. J. Paul, Dynamical solution to the problem of a small cosmological
constant and late-time cosmic acceleration, \textit{Physical Review Letters} \textbf{85} (2000) 4438.

\bibitem{ref8} I. Zlatev, W. Limin, and J. S. Paul, Quintessence, cosmic coincidence, and the cosmological constant, \textit{Physical
Review Letters} \textbf{82} (1999) 896.

\bibitem{ref9} P. Armendariz, V. M. Christian,  Mukhanov, and J. S. Paul. Essentials of k-essence, \textit{Physical Review D} \textbf{63} (2001)
103510.

\bibitem{ref10} H. \v{S}tefan\v{c}i\'{c}, Generalized phantom energy.\textit{Physics Letters B} \textbf{586} (2004) 5-10.

\bibitem{ref11} N. Bili\'{c}, B. Gary. Tupper, and R. D. Viollier. Unification of dark matter and dark energy: the inhomogeneous Chaplygin
gas. \textit{Physics Letters B} \textbf{535} (2002) 17-21.

\bibitem{ref12} S. Nojiri and S. D. Odintsov. Unified cosmic
history in modified gravity: from $F(R)$ theory to Lorentz non-invariant
models. \textit{Physics Reports} \textbf{505} (2011) 59-144.

\bibitem{ref13} S. Nojiri and D. Sergei. Introduction to
modified gravity and gravitational alternative for dark energy.\textit{International Journal of Geometric Methods in Modern Physics} \textbf{4} (2007) 115-145.

\bibitem{ref14} T. Harko et al. $f(R,T)$ gravity. \textit{Physical Review D}
\textbf{84} (2011) 024020.

\bibitem{ref15} M. Koussour, and M. Bennai. On a Bianchi type-I space-time
with bulk viscosity in $f(R,T)$ gravity. \textit{International Journal of Geometric
Methods in Modern Physics} (2021): 2250038.

\bibitem{ref16} M. Koussour, and M. Bennai. Cosmological models with
cubically varying deceleration parameter in $f(R,T)$ gravity. \textit{Afrika
Matematika} \textbf{33} (2022) 1-16.

\bibitem{ref17} M. Koussour et al. Holographic dark energy in
Gauss-Bonnet gravity with Granda-Oliveros cut-off. \textit{Nuclear Physics B}
(2022): 115738.

\bibitem{ref18} M. Koussour and M. Bennai. Stability analysis of
anisotropic Bianchi type-I cosmological model in teleparallel gravity. \textit{Classical and Quantum Gravity} \textbf{39} (2022) 105001.

\bibitem{ref19} J. Jim\'{e}nez, H. Lavinia and K. Tomi, Coincident general relativity. \textit{Physical Review D} \textbf{98} (2018)
044048.

\bibitem{ref20} S. Mandal, P. K. Sahoo, and J. R. L. Santos. \textit{Energy
conditions in $f(Q)$ gravity}. \textit{Physical Review D} \textbf{102} (2020) 024057.

\bibitem{ref21} S. Mandal, D. Wang, and P. K. Sahoo. Cosmography in $%
f(Q)$ gravity. \textit{Physical Review D} \textbf{102} (2020) 124029.

\bibitem{ref22} R. H. Ling and X. H. Zhai. Spherically symmetric
configuration in $f(Q)$ gravity. \textit{Physical Review D} \textbf{103} (2021) 124001.

\bibitem{ref23} N. Frusciante, Signatures of $f(Q)$ gravity in
cosmology. \textit{Physical Review D} \textbf{103} (2021) 044021.

\bibitem{ref24} W. Khyllep, P. Andronikos and J. Dutta. Cosmological solutions and growth index of matter
perturbations in $f(Q)$ gravity.\textit{ Physical Review D} \textbf{103} (2021) 103521.

\bibitem{ref25} T. Harko et al., Coupling matter in modified $Q$
gravity. \textit{Physical Review D} \textbf{98} (2018) 084043.

\bibitem{ref26} N. Dimakis, A. Paliathanasis, and T. Christodoulakis.
Quantum cosmology in $f(Q)$ theory. \textit{Classical and Quantum Gravity} \textbf{38} (2021) 225003.

\bibitem{ref27} B. Jing, T. H. Loo, and A. De. Geodesic
deviation equation in $f(Q)$ gravity. \textit{Chinese Journal of Physics} (2021).

\bibitem{ref28} De, Avik, et al. Isotropization of locally rotationally
symmetric Bianchi-I Universe in $f(Q)$ gravity. \textit{The European Physical
Journal C} \textbf{82} (2022) 1-11.

\bibitem{ref29} S. H. Shekh, Models of holographic dark energy in $f(Q)$
gravity. \textit{Physics of the Dark Universe} \textbf{33} (2021) 100850.

\bibitem{ref30} M. Koussour, S. H. Shekh, and M. Bennai. Cosmic
acceleration and energy conditions in symmetric teleparallel $f(Q)$
gravity. \textit{Journal of High Energy Astrophysics} \textbf{35} (2022) 43-51.

\bibitem{ref31} M. Koussour et al. Flat FLRW Universe in logarithmic
symmetric teleparallel gravity with observational constraints. \textit{Classical and Quantum Gravity} doi: 10.1088/1361-6382/ac8c7d (2022).

\bibitem{ref32} M. Koussour, S. H. Shekh, and M. Bennai. Anisotropic
nature of space--time in $f\left( Q\right) $ gravity. \textit{Physics of the Dark
Universe} (2022): 101051.

\bibitem{ref33} M. Koussour et al. Thermodynamical aspects of Bianchi
type-I Universe in quadratic form of $f(Q)$ gravity. \textit{arXiv preprint}
arXiv:2203.03639 (2022).

\bibitem{ref34} M. Koussour, S. H. Shekh, and M. Bennai. Bianchi type-I
Barrow holographic dark energy model in symmetric teleparallel gravity.
\textit{arXiv preprint} arXiv:2203.08181 (2022).

\bibitem{ref35} M. Koussour, S. H. Shekh, and M. Bennai. Anisotropic $f(Q)$
gravity model with bulk viscosity. \textit{arXiv preprint} arXiv:2203.10954 (2022).

\bibitem{ref36} G. Hooft, Dimensional reduction in quantum gravity.
\textit{arXiv preprint} gr-qc/9310026 (1993).

\bibitem{ref37} L. Susskind, The world as a hologram. \textit{Journal of
Mathematical Physics} \textbf{36} (1995) 6377-6396.

\bibitem{ref38} A. Cohen, B. David and A. E. Nelson.
Effective field theory, black holes, and the cosmological constant. \textit{
Physical Review Letters} \textbf{82} (1999) 4971.

\bibitem{ref39} M. Li, A model of holographic dark energy. \textit{Physics
Letters B} \textbf{603} (2004) 1-5.

\bibitem{ref40} J. Barrow, The area of a rough black hole. \textit{Physics
Letters B} \textbf{808} (2020) 135643.

\bibitem{ref41} E. N. Saridakis, Barrow holographic dark energy. \textit{Physical Review D} \textbf{102} (2020) 123525.

\bibitem{ref42} A. K. nagnostopoulos, S. Basilakos, and E. N. Saridakis. Observational constraints on Barrow holographic dark energy. \textit{The European Physical Journal C} \textbf{80} (2020) 1-9.

\bibitem{ref43} A. Priyanka, et al. Barrow holographic dark energy
in a nonflat Universe. \textit{Physical Review D} \textbf{104} (2021) 123519.

\bibitem{ref44} S. Srivastava and U. K. Sharma. Barrow
holographic dark energy with Hubble horizon as IR cutoff. \textit{International
Journal of Geometric Methods in Modern Physics} \textbf{18} (2021): 2150014.

\bibitem{ref45} U. K. Sharma, V. G. Varshney, and V. C. Dubey. Barrow agegraphic dark energy. \textit{International Journal of Modern
Physics D} \textbf{30} (2021) 2150021.

\bibitem{ref46} P. A. R. Ade et al. Planck 2015 results-XVI. Isotropy and
statistics of the CMB. \textit{Astronomy \& Astrophysics} \textbf{594} (2016) A16.

\bibitem{ref47} P. K. Sahoo, P. Sahoo, and B. K. Bishi. Anisotropic cosmological models in $f(R,T)$ gravity with variable
deceleration parameter. \textit{International Journal of Geometric Methods in
Modern Physics} \textbf{14} (2017) 1750097.

\bibitem{ref48} C. B. Collins and S. W. Hawking. Why is the
Universe isotropic?. \textit{The Astrophysical Journal} \textbf{180} (1973) 317-334.

\bibitem{Akarsu} O. Akarsu,  et al. Cosmology with hybrid
expansion law: scalar field reconstruction of cosmic history and
observational constraints. \textit{ Journal of Cosmology and Astroparticle Physics} \textbf{01} (2014) 022.

\bibitem{ref49} R. Solanki, Avik De, and P. K. Sahoo. Complete dark
energy scenario in $f(Q)$ gravity. \textit{Physics of the Dark Universe} (2022) 100996.

\bibitem{ref50} R. R. Caldwell, and E. V. Linder. Limits of
quintessence. \textit{Physical review letters} \textbf{95} (2005) 141301.

\bibitem{ref51} A. Nabila et al. Planck 2018 results-VI.
Cosmological parameters. \textit{Astronomy \& Astrophysics} \textbf{641} (2020) A6.

\bibitem{ref52} J. He and B. Wang. Effects of the interaction
between dark energy and dark matter on cosmological parameters. \textit{Journal of
Cosmology and Astroparticle Physics} \textbf{06} (2008) 010.

\bibitem{ref53} M. V. Santhi, and Y. Sobhanbabu. Bianchi type-III
Tsallis holographic dark energy model in Saez--Ballester theory of
gravitation. \textit{The European Physical Journal C} \textbf{80} (2020) 1-15.

\bibitem{ref54} B. Wang et al. Interacting dark energy and dark matter:
observational constraints from cosmological parameters. \textit{Nuclear Physics B} \textbf{778} (2007) 69-84.

\bibitem{ref55} S. Sarkar, Interacting holographic dark energy with
variable deceleration parameter and tachyon scalar field dark energy model
in LRS Bianchi type-II Universe. \textit{Astrophysics and Space Science} \textbf{350} (2014) 821-829.

\bibitem{Capozziello}  S. Capozziello and R. D'Agostino.
Model-independent reconstruction of $f(Q)$ non-metric gravity. \textit{Physics
Letters B} (2022): 137229.
\end{thebibliography}
\end{document}